\documentclass[%
 aip,
rsi,
 amsmath,amssymb,
reprint,%
numerical,%
floatfix,
]{revtex4-1}

\usepackage{graphicx}
\usepackage{dcolumn}
\usepackage{bm}

\usepackage[utf8]{inputenc}
\usepackage[T1]{fontenc}
\usepackage{mathptmx}

\begin{document}


\title[]{Setup for simultaneous electrochemical and color impedance measurements of electrochromic films: theory, assessment, and test measurement }

\author{Edgar A. Rojas-Gonz\'alez}
 \email{edgar.rojas@angstrom.uu.se.}
\author{Gunnar A. Niklasson}%

\affiliation{ 
Department of Engineering Sciences, The {\AA}ngstr{\"o}m Laboratory, Uppsala University, P.O. Box 534, SE-751 21 Uppsala, Sweden
}%


\date{\today}

\begin{abstract}
Combined frequency-resolved techniques are suitable to study electrochromic (EC) materials. We present an experimental setup for simultaneous electrochemical and color impedance studies of EC systems in transmission mode and estimate its frequency-dependent uncertainty by measuring the background noise. We define the frequency-dependent variables that are relevant to the combined measurement scheme, and a special emphasis is given to the complex optical capacitance and the complex differential coloration efficiency, which provide the relation between the electrical and optical responses. Results of a test measurement on amorphous $\mathrm{WO}_3$ with LED light sources of peak wavelengths of $470$, $530$, and $810~\mathrm{nm}$ are shown and discussed. In this case, the amplitude of the complex differential coloration efficiency presented a monotonous increase down to about $0.3~\mathrm{Hz}$ and was close to a constant value for lower frequencies. We study the effect of the excitation voltage amplitude on the linearity of the electrical and optical responses for the case of amorphous $\mathrm{WO}_3$ at $2.6~\mathrm{V}~\mathrm{vs.}~\mathrm{Li/Li}^+$, where a trade-off should be made between the signal-to-noise ratio (SNR) of the optical signal and the linearity of the system. For the studied case, it was possible to increase the upper accessible frequency of the combined techniques (defined in this work as the upper threshold of the frequency region for which the SNR of the optical signal is greater than $5$) from $11.2~\mathrm{Hz}$ to $125.9~\mathrm{Hz}$ while remaining in the linear regime with a tolerance of less than $5\%$.
\end{abstract}

\maketitle

\section{\label{sec1:level1}Introduction}
Around $40\%$ of the global energy use occurs in residential and commercial buildings, and approximately $65\%$ of this is generated by fossil fuel sources.\cite{Agency2017} Electrochromic (EC) smart windows can vary their optical transmission upon the application of an electrical voltage, which leads to an increase of comfort and a reduction of the need of active indoor climate regulation in a building.\cite{Granqvist2014} The optical absorption of EC materials is modified by the intercalation of ions from an electrolyte together with charge-balancing electrons from the outer circuit.\cite{Granqvist1995} It is actually the electrons that are responsible for the optical modulation since they populate previously empty states close to the Fermi level. Hence, the dynamics of the EC process involves the transport of both ions and electrons, being usually the ions the rate-limiting factor.

Frequency-resolved techniques such as electrochemical impedance spectroscopy (EIS) are able to probe the kinetic effects responsible for the electrical properties of EC materials and devices.\cite{Bonanos2018} The EIS uses a small voltage oscillation as an excitation, and measures the resulting oscillating current response. EIS allows to determine the kinetics of ion diffusion in EC materials by fitting the spectra to equivalent circuit models. It has been found that the diffusion process has fractal properties, so called anomalous diffusion,\cite{Malmgren2017} and effective diffusion coefficients can be determined from experimental data.

It has been hypothesized that the optical absorption process entails an additional step whereby a charge-compensating electron is trapped at an atomic color center.\cite{Bisquert2002,Bueno2008} Another spectroscopic technique, namely color impedance spectroscopy (CIS) is appropriate to study such effects in EC systems. The CIS technique is analogous to EIS with the difference that it measures the resulting oscillating optical response instead of the current. Simultaneous EIS and CIS measurements give a direct comparison between the optical variation of an EC material and the dynamics of the charges that induce the optical absorption. 

However, the realization of a combined EIS and CIS setup presents several experimental challenges in addition to the ones that are specifically related to EIS.\cite{Barsoukov2018} The relevant signals have to be synchronized and fed into the same frequency response analyzer (FRA). The light source needs to be stable throughout the duration of the experiment and the optical system conformed by the light source and the detector must present a good signal-to-noise ratio (SNR). The latter is necessary since the optical modulations associated with CIS are usually small because the linearity requirement restricts the magnitude of the excitation. The previous points, together with the inherent response of the sample of interest, determine the upper and lower bounds of the frequencies that are accessible in combined EIS and CIS measurements. 

The CIS technique was first used to study adsorption-desorption processes by frequency-modulated reflectance.\cite{Adzic1973} The technique was not used in optical transmittance mode, as far as we are aware, until around 1990.\cite{Kalaji1991} Previous works in which EC systems have been studied with a combination of EIS and CIS can be divided in two main categories. The first consists of consecutive (non-simultaneous) EIS and CIS in optical transmission mode.\cite{Bueno2008,Kalaji1991,Garcia-Belmonte2004,Amemiya1993,Amemiya1993a,Amemiya1994,Kim1997,Kim1997a} The second involves simultaneous EIS and CIS in optical reflectance mode,\cite{SusanaInesCordobaTorresi1990} and in a few cases also including mass impedance spectroscopy.\cite{Gabrielli1994,Agrisuelas2009}

The non-simultaneous case entails a risk of having variations in the studied sample between measurements and does not compare responses that come from a simultaneous excitation. The optical reflectance mode requires the EC material to be coated onto opaque metallic electrodes. Thus, it is not suitable for studying the combination of an EC material with a transparent conductive electrode, which is the case of many relevant applications{\textemdash}such as EC smart windows. Although some works have already used combined EIS and CIS measurements for obtaining relevant physical insights into EC systems, a thorough analysis of the performance and limitations of this approach is still lacking.

In this paper, we present an experimental setup for simultaneous EIS and CIS studies of EC systems in transmission mode at different optical wavelengths as well as its uncertainty estimation as a function of frequency. We focus our analysis on intercalation electrochromic systems, such as $\mathrm{WO}_3$, and we address the issue of improving the SNR at high frequencies by increasing the excitation voltage amplitude. The structure of the paper is described as follows. First, we define the relevant frequency-dependent quantities and elaborate the theoretical framework and concepts used in the analysis. Secondly, we describe the experimental setup and procedures. Thirdly, we present and discuss the performance assessment of the CIS setup and the results of a test measurement on amorphous $\mathrm{WO}_3$. Finally, we include some conclusions and remarks. 

\section{\label{sec2:level1}Theory}

During the frequency-resolved experiments, the time-dependent oscillatory excitation voltage $V(t)$ takes the form

\begin{equation}
V(t)=\langle V \rangle+V_\mathrm{A}  \sin(\omega t+\phi_V), \label{sec2:eq:1}
\end{equation}

with $\omega=2 \pi f$ the circular frequency associated with the linear frequency $f$,  $\langle V \rangle$ the stationary equilibrium bias, $V_\mathrm{A}$ the amplitude of the oscillation, and $\phi_V$ the reference phase{\textemdash}from now on, for simplicity, we set $\phi_V=0$. Similarly, assuming a linear system, the resulting time-dependent current $I(t)$, and transmittance $T(t)$ responses{\textemdash}with frequency-dependent amplitude and phase{\textemdash}can be expressed as 

\begin{eqnarray}
I(t)&=&\langle I \rangle+I_\mathrm{A}(\omega)  \sin[\omega t+\phi_I (\omega)],\label{sec2:eq:2}
\\
T(t)&=&\langle T \rangle+T_\mathrm{A}(\omega)  \sin[\omega t+\phi_\mathrm{op}(\omega)].\label{sec2:eq:3}
\end{eqnarray}

The amplitude $V_\mathrm{A}$ in Eq.~(\ref{sec2:eq:1}) is set by the conditions of the experiment, and the amplitudes and phases in Eqs.~(\ref{sec2:eq:2}) and (\ref{sec2:eq:3}) can be obtained experimentally by means of a FRA. A detailed theoretical description of the EIS technique and the working principles of frequency response analyzers can be found elsewhere.\cite{Barsoukov2018,Jonscher1996} In general, for each frequency, the FRA correlates each input with a reference signal of the form $\sin(\omega t)$  and its quadrature $\cos(\omega t)$. In the case of $I(t)$, this is done by performing the integrations

\begin{eqnarray}
(\omega/ N\pi) \int^{\delta+2\pi N/\omega}_{\delta} \mathrm{d}t I(t) \sin(\omega t) &=& I_\mathrm{A}(\omega) \cos{\phi_I{(\omega)}},\label{sec2:eq:4}\\
(\omega/ N\pi) \int^{\delta+2\pi N/\omega}_{\delta} \mathrm{d}t I(t) \cos(\omega t) &=& I_\mathrm{A}(\omega) \sin{\phi_I{(\omega)}},\label{sec2:eq:5}
\end{eqnarray}

with $N$ the integer number of cycles during which the integration takes place, and $\delta$ a time delay that is conveniently chosen for the stabilization of the responses upon the application of the sinusoidal excitation. The same applies to $T(t)$.

Then, using Euler's formula $\exp(i\phi)=\cos{\phi}+i\sin{\phi}$, it is possible to define the complex frequency-dependent current $\tilde{I}(\omega)$ and transmittance $\tilde{T}(\omega)$ as follows

\begin{eqnarray}
\tilde{I}(\omega)&=&I_\mathrm{A}(\omega)e^{i \phi_I{(\omega)}},\label{sec2:eq:6}\\
\tilde{T}(\omega)&=&T_\mathrm{A}(\omega)e^{i \phi_{\mathrm{op}}{(\omega)}}.\label{sec2:eq:7}
\end{eqnarray}

The relevant transfer functions can be obtained from different combinations of $V_\mathrm{A}$ and Eqs.~(\ref{sec2:eq:6}) and (\ref{sec2:eq:7}). Those are: the complex impedance

\begin{equation}
\tilde{Z}(\omega) \equiv V_\mathrm{A}/\tilde{I}(\omega)=|\tilde{Z}(\omega)|e^{i \phi_Z{(\omega)}}, \label{sec2:eq:8}
\end{equation}

with $\phi_Z(\omega) \equiv -\phi_I{(\omega)}$; the complex capacitance 

\begin{equation}
\tilde{C}(\omega) \equiv 1/[i\omega \tilde{Z}(\omega)]= \tilde{Q}(\omega)/V_\mathrm{A} =|\tilde{C}(\omega)|e^{i \phi_C{(\omega)}}, \label{sec2:eq:9}
\end{equation}

with $\phi_C{(\omega)} \equiv -\pi/2-\phi_Z{(\omega)}$ and $\tilde{Q}(\omega) \equiv (i\omega)^{-1} \tilde{I}(\omega)$ the complex frequency-dependent charge; a complex optical capacitance

\begin{equation}
\tilde{G}_{\mathrm{op}}(\omega) \equiv \langle T \rangle^{-1} [\tilde{T}(\omega)/V_\mathrm{A}]=|\tilde{G}_{\mathrm{op}}(\omega)|e^{i \phi_{\mathrm{op}}{(\omega)}}; \label{sec2:eq:10}
\end{equation}

and a complex differential coloration efficiency

\begin{eqnarray}
\tilde{K}(\omega) &\equiv& \tilde{G}_{\mathrm{op}}(\omega)/\tilde{C}(\omega)=  \langle T \rangle^{-1} [\tilde{T}(\omega)/\tilde{Q}(\omega)]\nonumber \\
&=&|\tilde{K}(\omega)|e^{i \phi_K{(\omega)}}, \label{sec2:eq:11}
\end{eqnarray}

with $\phi_K{(\omega)} \equiv \phi_{\mathrm{op}}{(\omega)}-\phi_I{(\omega)}+\pi/2$.

\begin{figure*}
\includegraphics[scale=0.5]{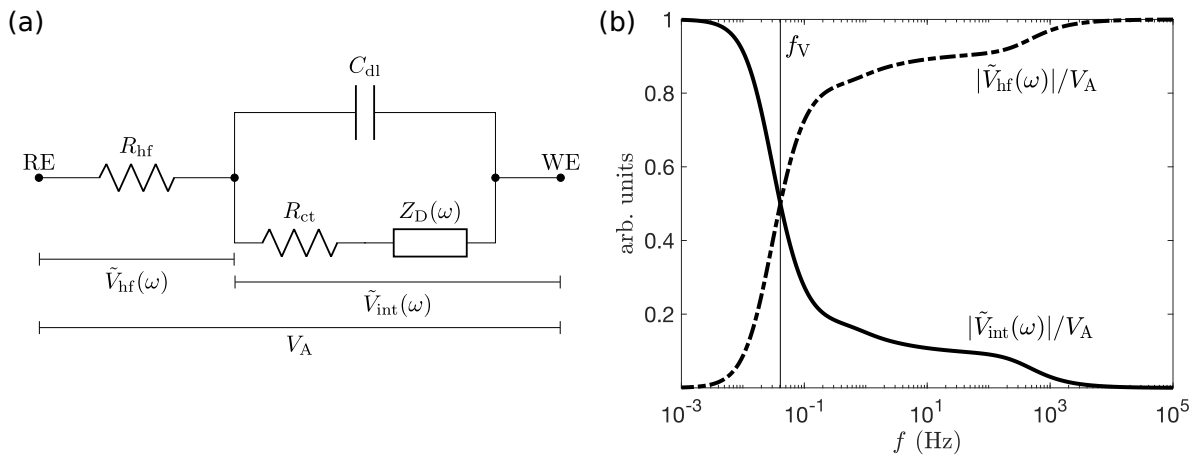}
\caption{\label{fig:1} (a) Equivalent circuit used to represent the qualitative features of the impedance spectra of a $\mathrm{WO}_3$ thin film electrode immersed in an electrolyte. The different elements are described in detail in the main text. (b) Simulation of the frequency dependence of the amplitudes of $\tilde{V}_{\mathrm{hf}}(\omega)$ and $\tilde{V}_{\mathrm{int}}(\omega)$ normalized with respect to $V_\mathrm{A}$. In (b), the vertical line depicts the transition frequency $f_\mathrm{V}$ and the curves were calculated using the equivalent circuit in (a) with parameters $R_\mathrm{hf}=100~\mathrm{\Omega}$, $C_\mathrm{dl}=50~\mathrm{\mu F}$, $R_\mathrm{ct}=10~\mathrm{\Omega}$, $R_\mathrm{D}=50~\mathrm{\Omega}$, $L=300~\mathrm{nm}$, and $D=5 \times 10^{-10}~\mathrm{cm}^2~\mathrm{s}^{-1}$.}
\end{figure*}

In the following, we comment about the relation between the amplitude of the excitation voltage and the linearity of intercalation systems, like $\mathrm{WO}_3$. A simplified equivalent circuit\cite{Ho1980} that describes the qualitative features of the impedance response of a $\mathrm{WO}_3$ thin film deposited onto a transparent conductive electrode and immersed in a lithium containing electrolyte is depicted in Fig.~\ref{fig:1}(a). Here, a high-frequency series resistance $R_\mathrm{hf}${\textemdash}due to the electrolyte and the transparent conductive electrode{\textemdash}is put in series with a combination consisting of a double layer capacitance $C_\mathrm{dl}${\textemdash}that develops at the interface between the electrode surface and the electrolyte{\textemdash}in parallel with an intercalation branch. The intercalation branch comprises a charge transfer resistance $R_\mathrm{ct}$ in series with a finite-space Warburg element defined as $Z_\mathrm{D}(\omega)=R_\mathrm{D}\coth{[(i\omega L^2/D)^{1/2}]}/(i\omega L^2/D)^{1/2}$, with $R_\mathrm{D}$ the diffusion resistance, $L$ the film thickness, and $D$ the chemical diffusion coefficient. 

The amplitude of the total applied oscillating voltage $V_\mathrm{A}$ is measured between the working (WE) and the reference (RE) electrodes, whereas the complex voltage drops across the high-frequency resistance, and between the terminals of the intercalation branch are denoted by $\tilde{V}_{\mathrm{hf}}(\omega)$, and $\tilde{V}_{\mathrm{int}}(\omega)$, respectively. For descriptive purposes, the qualitative frequency dependence of $\tilde{V}_{\mathrm{hf}}(\omega)$ and $\tilde{V}_{\mathrm{int}}(\omega)$ is depicted in Fig.~\ref{fig:1}(b) using illustrative values of the circuit elements of Fig.~\ref{fig:1}(a). 

Figure~\ref{fig:1}(b) outlines a phenomenon explained in detail by Darowicki in Ref.~\onlinecite{Darowicki1997} where the author studies an electrode presenting parallel Faradaic and non-Faradaic processes in series with a resistance assigned to an ohmic drop. That is, the effective amplitude of the sinusoidal potential at the electrode interface $|\tilde{V}_{\mathrm{int}}(\omega)|$ differs from the total applied one $V_\mathrm{A}$ and this discrepancy varies with frequency. In fact, at high frequencies most of the applied potential is dropped at $R_\mathrm{hf}$, whereas $|\tilde{V}_{\mathrm{int}}(\omega)|$ becomes similar to $V_\mathrm{A}$ at low frequencies, and the transition between these two regions occurs around a certain threshold{\textemdash}denoted by $f_\mathrm{V}$ in Fig.~\ref{fig:1}(b).

The coloration in an electrochromic electrode is assigned to the intercalation branch. Thus, in principle, we can aim for high values of $|\tilde{V}_{\mathrm{int}}(\omega)|$ to increase the SNR of the optical signal. Note that, for example, a constant target value of $|\tilde{V}_{\mathrm{int}}(\omega)|$ throughout the whole frequency range would require considerable higher values of the total applied voltage at the high frequencies with respect to those at low frequencies. The upper bound of $V_\mathrm{A}$ is determined by the linearity condition of the electrochemical system, which is only regulated by the parallel combination in Fig.~\ref{fig:1}(a) because $R_{\mathrm{hf}}$ is intrinsically linear. In addition, the strong voltage dependence of $R_{\mathrm{ct}}$, $Z_\mathrm{D}(\omega)$, and $C_{\mathrm{dl}}${\textemdash}that leads to a non-linear behavior{\textemdash}is mentioned by Ho, Raistrick, and Huggins in Ref.~\onlinecite{Ho1980} and shown experimentally elsewhere\cite{Mattsson2000,Stromme1996} (note that in the latter works the double layer capacitance is modeled by a constant phase element instead of using a pure capacitor). As a result, for a constant value of $V_\mathrm{A}$, the non-linear effects are expected to be more relevant at low frequencies.

In the context of this work, the previous arguments justify the employment of a variable-amplitude method\cite{Hirschorn2008} for improving the SNR of the optical signal while staying in the linear regime. Indeed, we can apply high, and low excitation voltage amplitudes at high, and low frequencies, respectively. A theoretical determination of the appropriate voltage amplitudes for the different frequency ranges of interest would require prior knowledge of the characteristics of the system and is out of the scope of this work. Instead, an experimental approach can be used. For example, an EIS spectrum employing a small excitation voltage amplitude can be measured. Then, subsequent spectra can be taken with increasing voltage amplitudes until a good SNR is obtained in the CIS spectrum. This can be done while the EIS spectra remain similar to that of the low-amplitude case up to a certain chosen tolerance{\textemdash}at least for a given portion of the high frequency region of the spectrum, which is the one that usually requires a larger improvement in terms of the SNR of the optical signal.

\section{\label{sec3:level1} Experimental setup and procedures}

\subsection{\label{sec3:level2_1}Electrode preparation and electrochemical cell}

The amorphous tungsten oxide ($\mathrm{WO}_3$) films used in this work were prepared by reactive DC magnetron sputtering in a Balzers UTT 400 unit. A $99.95\%$ pure metallic W target was used in a $99.995\%$ pure $\mathrm{O}_2/\mathrm{Ar}$ atmosphere at $30~\mathrm{mTorr}$. The deposition was performed at a discharge power of 240 W, Ar flow of $50~\mathrm{ml/min}$, and $\mathrm{O}_2$ flow of $22~\mathrm{ml/min}$. The films ($\sim300~\mathrm{nm}$ thick) were deposited onto an unheated glass substrate pre-coated with conducting $\mathrm{In}_2\mathrm{O}_3$:$\mathrm{Sn}$ (ITO) having a sheet resistance of $15~\Omega/\mathrm{sq}$. The film thicknesses were determined by stylus profilometry using a Bruker DektakXT instrument. The amorphous structure of the $\mathrm{WO}_3$ was confirmed by X-ray diffraction (XRD) patterns collected with a Siemens D5000 diffractometer using Cu $K\alpha$ radiation.
A standard three-electrode setup was used for the electrochemical measurements in an argon-filled glove box ($\mathrm{H}_2\mathrm{O}~\mathrm{level}< 0.6~\mathrm{ppm}$). A quartz cuvette was filled with the electrolyte, which consisted of $1\mathrm{M}~\mathrm{LiClO}_4$ dissolved in propylene carbonate. Unless specified, a $\mathrm{WO}_3$ film acted as the WE, and lithium foils were used both as the counter electrode (CE) and the RE. Throughout this work, the active area of the WE was $\sim1~\mathrm{cm}^2$.

\subsection{\label{sec3:level2_2}Electrochemical techniques, and combined EIS and CIS experimental setup}

\begin{figure}
\includegraphics[scale=0.65]{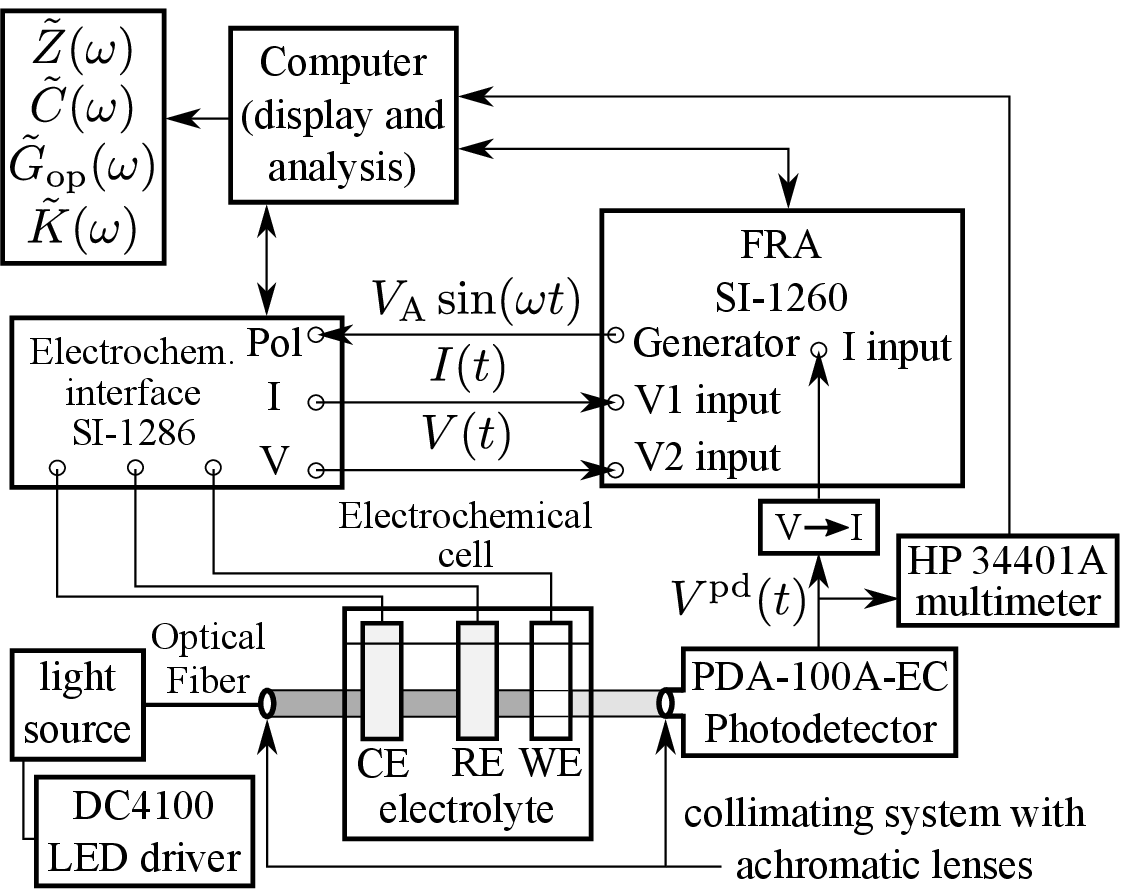}
\caption{\label{fig:2} Schematics of the combined EIS and CIS experimental setup.}
\end{figure}

The combined EIS and CIS experimental setup is depicted in Fig.~\ref{fig:2}. The Cyclic voltammetry (CV) and potentiostatic polarization techniques were performed by an electrochemical interface (SI-1286, Solartron), which was connected to the electrodes in the electrochemical cell. It measured the relative potential between the WE and the RE, and the current flowing between the WE and the CE. The SI-1286 also controlled the relative potential of the WE when it was required. A frequency response analyzer (SI-1260, Solartron) was used in combination with the electrochemical interface during the simultaneous EIS and CIS measurements. In this case, the FRA supplied a sinusoidal excitation signal to the SI-1286, and the values measured by the SI-1286 were feed to the FRA inputs.

The light source consisted of a fiber-coupled LED controlled in constant current mode by a LED driver (DC4100, Thorlabs). We used LEDs of three different peak wavelengths in the present work{\textemdash}namely, $810~\mathrm{nm}$ (M810F2, Thorlabs), $530~\mathrm{nm}$ (M530F2, Thorlabs), and $470~\mathrm{nm}$ (M470F3, Thorlabs). An optical fiber ($400~\mathrm{\mu m}$ in diameter, $0.39~\mathrm{NA}$; M28L01, Thorlabs) was connected to a collimating system (74-ACH, Ocean Optics) with achromatic lenses (74-ACR, Ocean Optics). Unless specified, the collimated beam of light passed through the quartz cuvette, the electrolyte, and the WE before being collected by a photodetector (PDA-100A-EC, Thorlabs). During the frequency-resolved measurements the output signal of the photodetector $V^{\mathrm{pd}}(t)$ takes the form

\begin{equation}
V^{\mathrm{pd}}(t) = \langle V^{\mathrm{pd}} \rangle + V^{\mathrm{pd}}_\mathrm{A}(\omega) \sin[\omega t + \phi_{\mathrm{op}}(\omega)],\label{sec2:eq:12}
\end{equation}

with $\langle V^{\mathrm{pd}} \rangle$  its stationary equilibrium bias, $V^{\mathrm{pd}}_\mathrm{A}(\omega)$  its frequency-dependent amplitude response, and $\phi_{\mathrm{op}}(\omega)$ the same phase as in Eq.~(\ref{sec2:eq:3}). Then, we can define a complex frequency-dependent photodetector voltage as $\tilde{V}^{\mathrm{pd}}(\omega) \equiv V^{\mathrm{pd}}_\mathrm{A}(\omega) \exp[i\phi_{\mathrm{op}}(\omega)]$. It is important to remark that the FRA does not give information about the total  value of $V^{\mathrm{pd}}(t) $ because it only measures the optical signal at the exciting frequency.  Thus, $V^{\mathrm{pd}}(t) $ was simultaneously monitored by a digital multimeter (34401A, HP/Agilent) and recorded by a computer during the whole experimental sequence in order to obtain the value of $\langle V^{\mathrm{pd}} \rangle$. 

Following a similar approach to the one presented by Singh and Richert in Ref.~\onlinecite{Singh2012}, the $V^{\mathrm{pd}}(t)$ was fed into the current input of the frequency response analyzer via a $150~\Omega$ resistor. Previously, we had performed a calibration of the FRA output data related to the FRA current input for retrieving the proper values of  $\tilde{V}^{\mathrm{pd}}(\omega)$. We defined the transmittance in Eq.~(\ref{sec2:eq:3}) as

\begin{equation}
T(t)=V^{\mathrm{pd}}(t)/V^{\mathrm{pd}}_\mathrm{B},\label{sec2:eq:13}
\end{equation}

where the output value of the photodetector at the bleached state $V^{\mathrm{pd}}_\mathrm{B}$ (that is, for $\mathrm{WO}_3$ at $4.0~\mathrm{V}~\mathrm{vs.}~\mathrm{Li}/\mathrm{Li}^+$) was chosen as the $100\%$ transmittance reference. Then, $\langle T \rangle = \langle V^{\mathrm{pd}} \rangle/V^{\mathrm{pd}}_\mathrm{B}$, and $\tilde{T}(\omega)=\tilde{V}^{\mathrm{pd}}(\omega)/V^{\mathrm{pd}}_\mathrm{B}$.  In this work, $V^{\mathrm{pd}}_\mathrm{B}$ presented typical values between $6$ and $8~\mathrm{V}$.

Finally, the real and complex components of the frequency-dependent variables measured by the FRA were recorded by the computer and the transfer functions defined in Eqs.~(\ref{sec2:eq:8})-(\ref{sec2:eq:11}) were calculated and stored for further analysis.

Also, a preliminary CV measurement was done on a $\mathrm{WO}_3$ WE at $10~\mathrm{mV}/\mathrm{s}$ using a BioLogic SP-200 potentiostat. Here, the current density and the transmittance (for the $530~\mathrm{nm}$ LED) were simultaneously measured by feeding the photodetector output signal into an analog input of the SP-200 potentiostat.

\subsection{\label{sec3:level2_3}Simultaneous EIS and CIS measurements}

In the following, we explain the experimental procedure for the background noise assessment, the test measurement on amorphous $\mathrm{WO}_3$, and the variable-amplitude method aimed to increase the SNR of the optical signal{\textemdash}defined here as the ratio between the experimental value of $|\tilde{G}_{\mathrm{op}}(\omega)|$ and its respective standard deviation. It is important to mention that, before each experimental sequence, the selected LED was let to stabilize for at least $30~\mathrm{min}$ at the desired intensity.

We assessed the background noise level of the CIS using an uncoated ITO electrode (with sheet resistance of $15~\mathrm{\Omega}/\mathrm{sq}$) as the WE. In this case, the WE was placed outside the optical path{\textemdash}that is, the beam of light passed only through the quartz cuvette and the electrolyte. The WE presented an open circuit potential of around $3.3~\mathrm{V}~\mathrm{vs.}~\mathrm{Li}/\mathrm{Li}^+$ when it was first immersed in the electrolyte. For each optical wavelength, the experimental sequence comprised an initial CV during three cycles in the voltage range of $2.7-3.7~\mathrm{V}~\mathrm{vs.}~\mathrm{Li}/\mathrm{Li}^+$ at $50~\mathrm{mV}/\mathrm{s}$, followed by a potentiostatic polarization treatment at $3.2~\mathrm{V}~\mathrm{vs.}~\mathrm{Li}/\mathrm{Li}^+$ for $30$ minutes, and $10$ sequential simultaneous EIS and CIS measurements at $3.2~\mathrm{V}~\mathrm{vs.}~\mathrm{Li}/\mathrm{Li}^+$ in the frequency range between $10~\mathrm{mHz}$ and $30~\mathrm{kHz}$ for an excitation voltage of $20~\mathrm{mV}$ root-mean-square (rms). The mean and the unbiased estimation of the standard deviation of the real and imaginary components of $\tilde{V}^{\mathrm{pd}}(\omega)$ were calculated for each set of $10$ measurements. The standard deviations calculated here were used as the uncertainty estimation for further measurements related to the optical signal. When it was required, the propagation of the uncertainty was performed assuming a gaussian distribution of the errors. In this work, we assumed that the uncertainties assigned to the electrochemical measurements were negligible with respect to those of the optical signal.

For each optical wavelength, the test measurement on a $\mathrm{WO}_3$ electrode consisted of the experimental sequence described as follows. Initially, a CV was performed in the voltage range of $2.0-4.0~\mathrm{V}~\mathrm{vs.}~\mathrm{Li}/\mathrm{Li}^+$ at a rate of $5~\mathrm{mV}/\mathrm{s}$ during three cycles. Next, a potentiostatic polarization treatment was carried out at the equilibrium potential of interest during $20~\mathrm{min}$, ($2.6~\mathrm{V}~\mathrm{vs.}~\mathrm{Li}/\mathrm{Li}^+$ in this work), which let the WE reach its electrochemical steady-state condition. Finally, the frequency-resolved measurements{\textemdash}with an integration during $4$ cycles after a delay of $1$ cycle for each measured frequency{\textemdash}were done in the frequency range between $10~\mathrm{mHz}$ and $30~\mathrm{kHz}$ using an excitation voltage amplitude of $20~\mathrm{mV}~\mathrm{rms}$. 

The variable-amplitude method was carried out using the $810~\mathrm{nm}$ LED and performing simultaneous EIS and CIS measurements on a $\mathrm{WO}_3$ WE at a bias of $2.6~\mathrm{V}~\mathrm{vs.}~\mathrm{Li}/\mathrm{Li}^+$. In this case, we first measured a CV during three cycles in the voltage range of $2.0-4.0~\mathrm{V}~\mathrm{vs.}~\mathrm{Li}/\mathrm{Li}^+$ at $5~\mathrm{mV}/\mathrm{s}$. Then, a potentiostatic polarization treatment was done at $2.6~\mathrm{V}~\mathrm{vs.}~\mathrm{Li}/\mathrm{Li}^+$ for $20$ minutes. Finally, we performed consecutive simultaneous EIS and CIS measurements at a bias of $2.6~\mathrm{V}~\mathrm{vs.}~\mathrm{Li}/\mathrm{Li}^+$ in the frequency range between $10~\mathrm{mHz}$ and $30~\mathrm{kHz}$ for excitation voltage amplitude values ranging from $10$ to $500~\mathrm{mV}~\mathrm{rms}${\textemdash}with transitions consisting of potentiostatic polarization treatments at $2.6~\mathrm{V}~\mathrm{vs.}~\mathrm{Li}/\mathrm{Li}^+$ during $10$ minutes.

\section{\label{sec4:level1}Results and discussion}

\begin{figure}[h]
\includegraphics[scale=0.49]{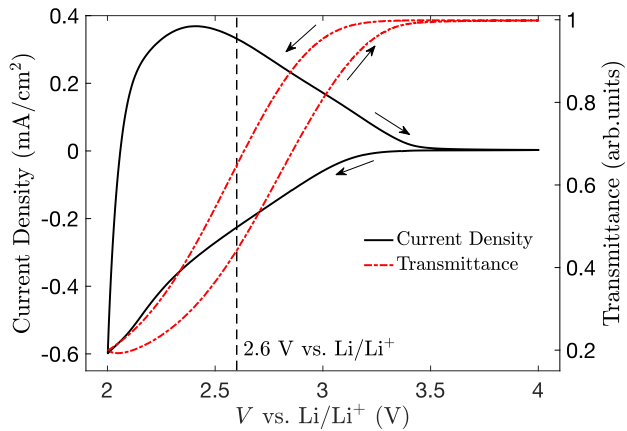}
\caption{\label{fig:3} Cyclic voltammogram for an amorphous $\mathrm{WO}_3$ film on ITO ($15~\mathrm{\Omega/sq}$) coated glass at $10~\mathrm{mV}/\mathrm{s}$. The current density (solid black curve) and the simultaneously measured transmittance using the $530~\mathrm{nm}$ LED (red dashed curve) are shown. The arrows indicate the sweeping direction. The vertical dashed line depicts the position of the $2.6~\mathrm{V}~\mathrm{vs.}~\mathrm{Li}/\mathrm{Li}^+$ WE potential.}
\end{figure}

The cyclic voltammogram of an amorphous $\mathrm{WO}_3$ WE and its simultaneously measured transmittance suggest linear regions around $2.6~\mathrm{V}~\mathrm{vs.}~\mathrm{Li}/\mathrm{Li}^+$, see Fig.~\ref{fig:3}{\textemdash}especially if we look at the intercalation branches (the ones denoted by the arrows pointing toward the left-hand side). It is also observed that the transmittance exhibits a more extended linear region than the corresponding current response. Thus, we chose this bias potential for the test measurement and the variable-amplitude study.

\subsection{\label{sec4:level2_1}Background noise determination}

\begin{figure}[t!]
\includegraphics[scale=0.50]{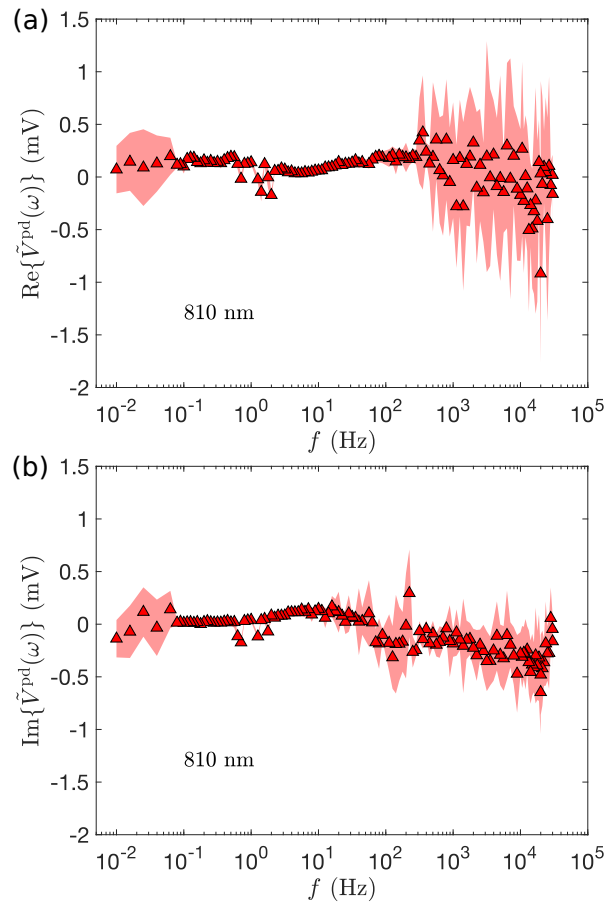}
\caption{\label{fig:4} Background noise of the real (a) and complex (b) components of $\tilde{V}^{\mathrm{pd}}(\omega)$ as a function of frequency for the $810~\mathrm{nm}$ LED. The symbols represent the mean values of $10$ measurements and the shaded regions show one standard deviation around the mean values.}
\end{figure}

The background noise levels for the real and imaginary components of $\tilde{V}^{\mathrm{pd}}(\omega)$ for the $810~\mathrm{nm}$ LED are depicted in Fig.~\ref{fig:4}. The results for the other light sources, not shown here, are almost identical. As expected from random noise, the mean values are close to zero. The order of magnitude of the vertical axes in Fig.~\ref{fig:4} is better understood by considering that $V^{\mathrm{pd}}_\mathrm{B}\sim7~\mathrm{V}$ under typical conditions. Then, a change of $\sim1~\mathrm{mV}$ in the photodetector output signal would correspond to a transmittance variation of $\sim0.01\%$. 

A close look to the frequency dependence of the standard deviations in Fig.~\ref{fig:4} shows three characteristics sections. The first region corresponds to the high frequency part of the spectrum and presents a relatively high noise level that may originate from electrical interferences{\textemdash}it extends down to $\sim100~\mathrm{Hz}$, and $\sim10~\mathrm{Hz}$ in Figs.~\ref{fig:4}(a), and \ref{fig:4}(b), respectively. This is the most problematic in terms of CIS because, in this frequency range, the optical response is often of the order or smaller than the noise level. The second region exhibits a nearly negligible noise level down to $\sim0.1~\mathrm{Hz}$ and is well-suited for CIS measurements. The third region is located below $\sim0.1~\mathrm{Hz}$ and shows a relatively high noise level that may arise from very small drifts in the optical stationary equilibrium bias $\langle V^{\mathrm{pd}} \rangle$. Generally, this should not represent an issue because the SNR is usually high within this frequency range.

\subsection{\label{sec4:level2_2}Test measurement on amorphous $\mathrm{WO}_3$}

\begin{figure*}[h!]
\includegraphics[scale=0.50]{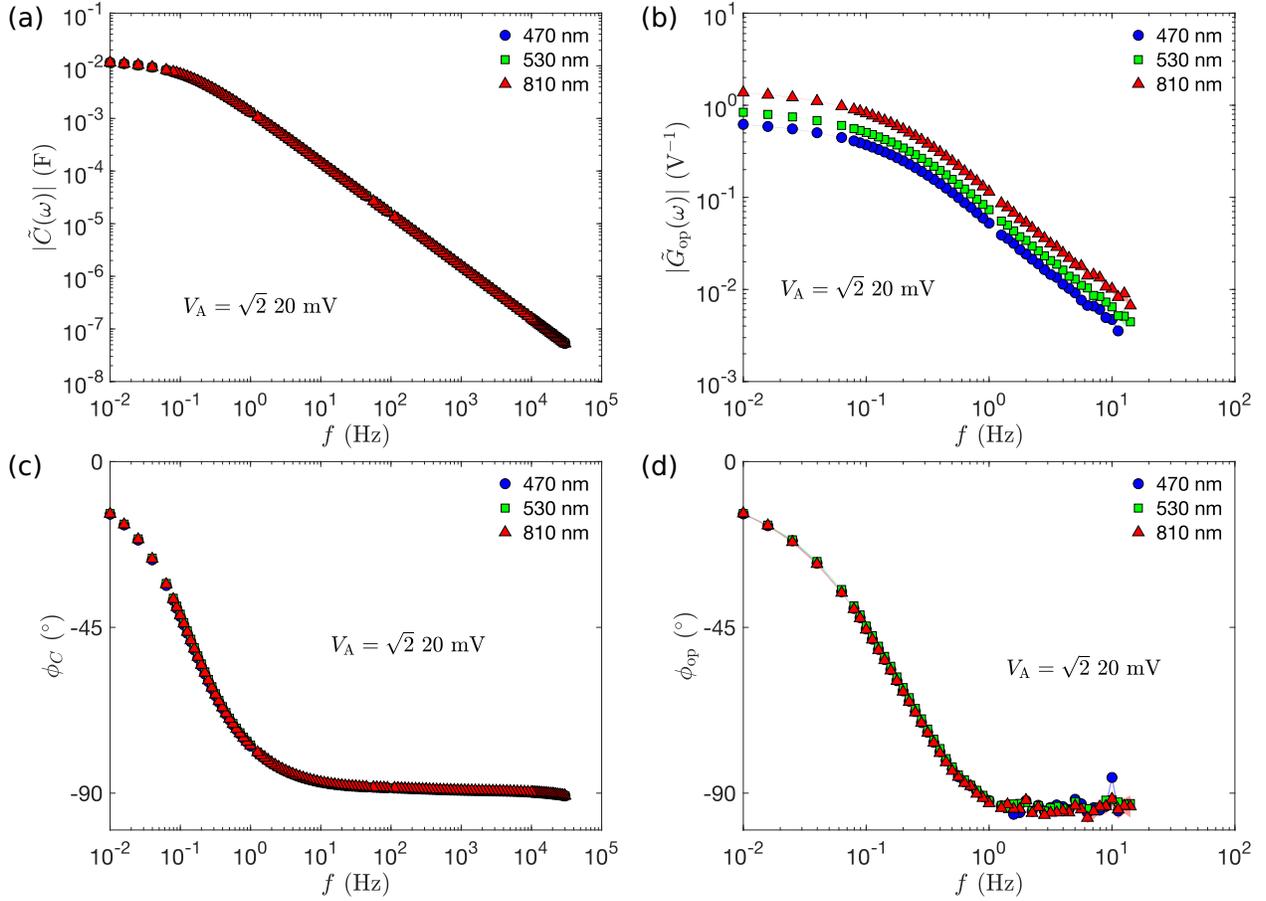}
\caption{\label{fig:5} Measured amplitude (a), and phase (c) of $\tilde{C}(\omega)$, and amplitude (b) and phase (d) of $\tilde{G}_{\mathrm{op}}(\omega)$ for the $470~\mathrm{nm}$ (blue circles), $530~\mathrm{nm}$ (green squares), and $810~\mathrm{nm}$ (red triangles) LEDs. The excitation voltage amplitude was $20~\mathrm{mV}~\mathrm{rms}$. The shaded regions in (b) and (d){\textemdash}small compared to the scale of the symbols{\textemdash}show one standard deviation around the experimental values. In (b) and (d), each data set is truncated so that the high-frequency region with SNR of $|\tilde{G}_{\mathrm{op}}(\omega)|$ smaller than $5$ is not shown. }
\end{figure*}

The measured complex capacitance and the complex optical capacitance of amorphous $\mathrm{WO}_3$ at $2.6~\mathrm{V}~\mathrm{vs.}~\mathrm{Li}/\mathrm{Li}^+$ are depicted in Fig.~\ref{fig:5}.

As expected, there is a resemblance between their respective absolute values and phases. Besides, the value of $|\tilde{G}_{\mathrm{op}}(\omega)|$ is higher at a wavelength of $810~\mathrm{nm}$, and lower at $470~\mathrm{nm}$, see Fig.~\ref{fig:5}(b). This is in accordance with the well-known  increase of the coloration efficiency of amorphous $\mathrm{WO}_3$ toward the infrared.\cite{Berggren2001,Triana2015} On the other hand, there are no noticeable discrepancies in $\phi_{\mathrm{op}}(\omega)$ for the different optical wavelengths, see Fig.~\ref{fig:5}(d). In addition, the measured complex capacitance is almost identical for the given optical wavelengths, see Figs.~\ref{fig:5}(a) and \ref{fig:5}(c), which shows that the electrochemical system did not vary appreciably between experiments.

Three sections can be observed in Fig.~\ref{fig:5}. The first region, for frequencies higher than $\sim1~\mathrm{Hz}$, presents phases close to $-90^\circ$, which is characteristic of a system with a predominant resistive response. The second region, for frequencies between $\sim0.1~\mathrm{Hz}$  and  $\sim1~\mathrm{Hz}$, outlines a transition toward a dominant capacitive response. The latter is shown in the third region for frequencies below $\sim0.1~\mathrm{Hz}$.

\begin{figure}[b!]
\includegraphics[scale=0.50]{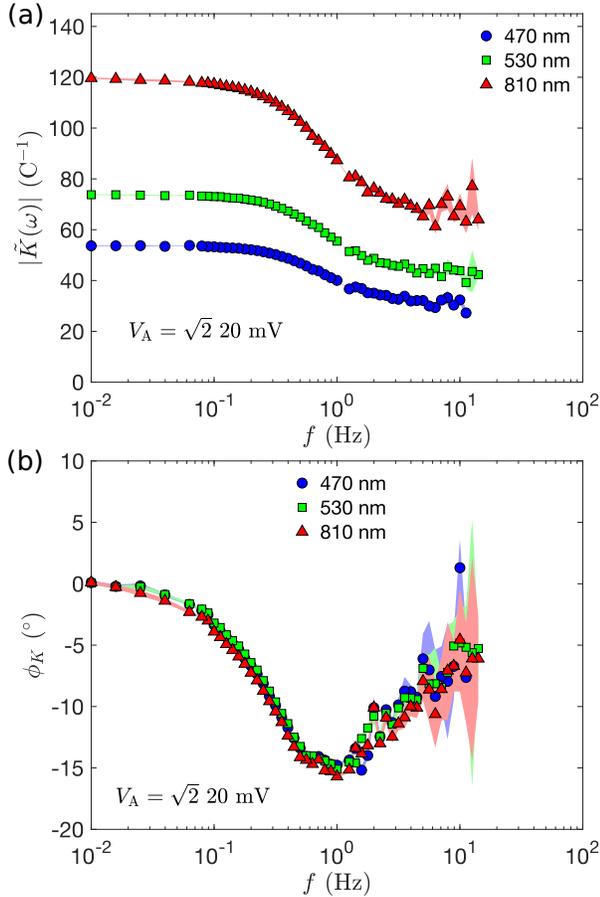}
\caption{\label{fig:6} Amplitude (a) and phase (b) of $\tilde{K}(\omega)$ calculated from the data in Fig.~\ref{fig:5}. Light source wavelengths of $470~\mathrm{nm}$ (blue circles), $530~\mathrm{nm}$ (green squares), and $810~\mathrm{nm}$ (red triangles) are depicted. The shaded regions show one standard deviation around the experimental values. Each data set is truncated so that the high-frequency region with SNR of $|\tilde{G}_{\mathrm{op}}(\omega)|$ smaller than $5$ is not shown. }
\end{figure}

The complex differential coloration efficiency obtained from the data presented in Fig.~\ref{fig:5} was calculated using Eq.~(\ref{sec2:eq:11}) and the results are depicted in Fig.~\ref{fig:6}. 

$\tilde{K}(\omega)$ is arguably the most interesting quantity that can be obtained from a combined EIS and CIS setup. Its absolute value, see Fig.~\ref{fig:6}(a), gives the frequency-dependent amplitude of the optical modulations per unit charge. Similar to $|\tilde{G}_{\mathrm{op}}(\omega)|$, $|\tilde{K}(\omega)|$ is larger for a wavelength of $810~\mathrm{nm}$ and smaller for $470~\mathrm{nm}$. Moreover, it increases steadily down to $\sim0.3~\mathrm{Hz}$ and reaches a plateau-like region for smaller frequencies. 

A delay between the optical variations and the charges can be clearly seen in Fig.~\ref{fig:6}(b){\textemdash}that is, a departure from $0^\circ$. Its maximum occurs at $\sim1~\mathrm{Hz}$, while it tends to decrease toward the low and high frequencies. Furthermore, no significant differences between the studied optical wavelengths can be observed in Fig.~\ref{fig:6}(b). The simultaneity of the EIS and CIS measurements allows to obtain $\phi_K(\omega)$ in a reliable and accurate way because it does not depend on a calculation that uses results from independent measurements.

\subsection{\label{sec4:level2_3}Variable-amplitude method}

\begin{figure*}[hb]
\includegraphics[scale=0.50]{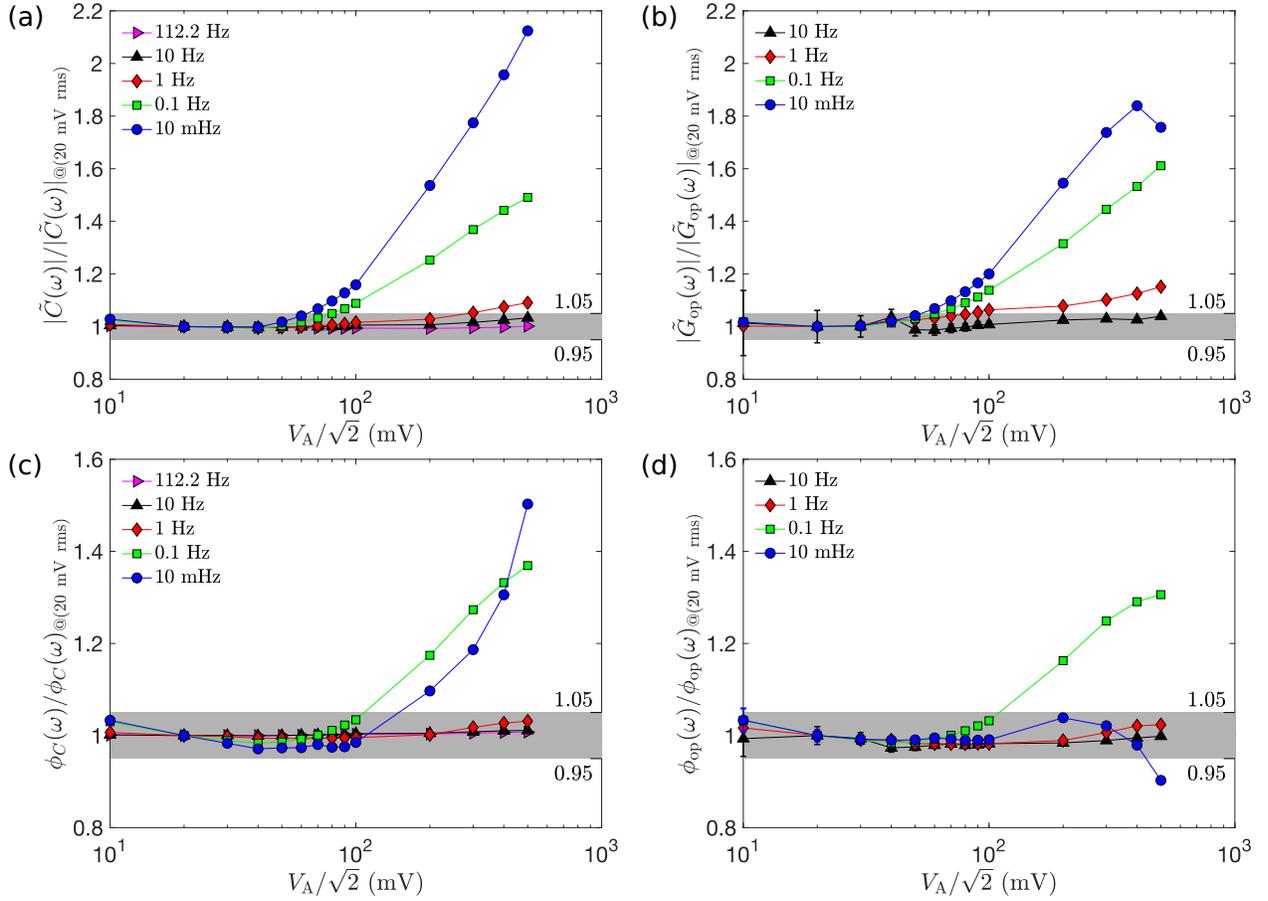}
\caption{\label{fig:7} Absolute value (a) and phase (c) of $\tilde{C}(\omega)$, and absolute value (b) and phase (d) of $\tilde{G}_{\mathrm{op}}(\omega)$ as a function of excitation voltage amplitude for $10~\mathrm{Hz}$ (black up-pointing triangles), $1~\mathrm{Hz}$ (red diamonds), $0.1~\mathrm{Hz}$ (green squares), and $10~\mathrm{mHz}$ (blue circles). An additional data set for $112.2~\mathrm{Hz}$ (magenta right-pointing triangles) is depicted in (a) and (c). Each quantity is normalized with respect to its value at $20~\mathrm{mV}~\mathrm{rms}$, and the error bars in (b) and (d){\textemdash}smaller than the symbols for some data points{\textemdash}show one standard deviation around the experimental values. The shadowed regions depict a $5\%$ tolerance with respect to the values at $20~\mathrm{mV}~\mathrm{rms}$. The symbols, connected by straight lines, denote the experimental data. }
\end{figure*}

Here, we explore the possibility of improving the SNR of the optical signal at high frequencies. This can be readily achieved by increasing the excitation voltage amplitude. However, the effect of such procedure on the departure from the linearity condition must be studied and quantified. In the linear regime, neither the complex capacitance nor the complex optical capacitance should depend on the excitation voltage amplitude, the contrary is an indication that the system is in a non-linear regime. The effect of the variation of the excitation voltage amplitude on the complex capacitance and the complex optical capacitance is depicted in Fig.~\ref{fig:7}. 

We chose $20~\mathrm{mV}~\mathrm{rms}$ as the reference excitation voltage amplitude and defined the linear regime by setting a tolerance of $5\%$ with respect to the experimental values at the given reference excitation voltage amplitude. In Fig.~\ref{fig:7}, the linear regime ranges are portrayed by shaded regions. 

Figure~\ref{fig:7} shows that the lower frequencies deviate rapidly from the linear regime, whereas the higher the frequency the less is the perturbation upon the increase of the excitation voltage amplitude. These results are in accordance with the qualitative analysis that we elaborated around the simplified model presented in Fig.~\ref{fig:1}. An important consequence from the results depicted in Fig.~\ref{fig:7} is that, provided that we accept the $5\%$ tolerance criterion, we could choose to use an excitation voltage amplitude up to $100$, and $500~\mathrm{mV}~\mathrm{rms}$ for frequencies higher than $1$, and $10~\mathrm{Hz}$, respectively{\textemdash}these limits were set by $|\tilde{G}_{\mathrm{op}}(\omega)|$ because it is the quantity that departs from the linear regime at a lower excitation voltage amplitude, this can be seen by comparing for example the curve corresponding to $1~\mathrm{Hz}$ in Fig.~\ref{fig:7}(b) to those in Figs.~\ref{fig:7}(a), \ref{fig:7}(c), and \ref{fig:7}(d). By doing so, the uncertainties at high frequencies could be drastically diminished{\textemdash}which is precisely the region of the spectrum that is problematic for the $\mathrm{WO}_3$ case, see Fig.~\ref{fig:6}.

\begin{figure*}
\includegraphics[scale=0.50]{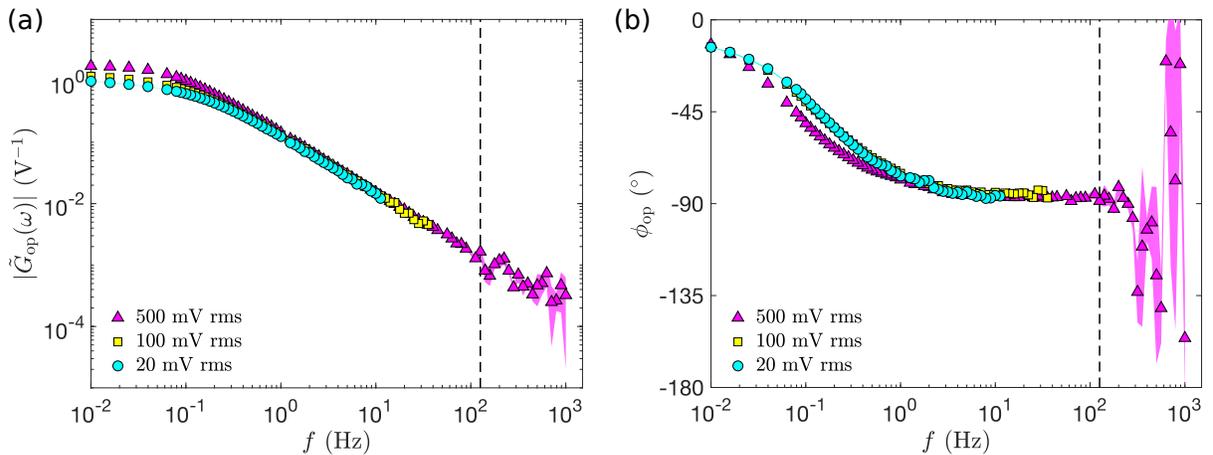}
\caption{\label{fig:8} Measured amplitude (a), and phase (b) of $\tilde{G}_{\mathrm{op}}(\omega)$ for excitation voltage amplitudes of $20~\mathrm{mV}~\mathrm{rms}$ (cyan circles), $100~\mathrm{mV}~\mathrm{rms}$ (yellow squares), and $500~\mathrm{mV}~\mathrm{rms}$ (magenta triangles). The shaded regions show one standard deviation around the experimental values. On the left-hand side of the vertical dashed line, only the frequency region with SNR of $|\tilde{G}_{\mathrm{op}}(\omega)|$ greater than $5$ is shown for each data set. The data for $500~\mathrm{mV}~\mathrm{rms}$ located on the right-hand side of the vertical dashed line present SNR of $|\tilde{G}_{\mathrm{op}}(\omega)|$ smaller than $5$. }
\end{figure*}

Indeed, Fig.~\ref{fig:8} shows that for the $\mathrm{WO}_3$ WE case the upper accessible frequencies corresponding to excitation voltage amplitudes of $20$, $100$ and $500~\mathrm{mV}~\mathrm{rms}$ are $11.2$, $35.5$, and $125.9~\mathrm{Hz}$, respectively. 

In this work, the upper accessible frequency was defined as the upper threshold of the frequency region for which the of SNR of $|\tilde{G}_{\mathrm{op}}(\omega)|$ is greater than $5$. The data on the right-hand side of the vertical dashed lines in Fig.~\ref{fig:8} are located beyond the upper accessible frequency for the $500~\mathrm{mV}~\mathrm{rms}$ case and were included in the plots to depict the transition toward a noisy region with high uncertainty values. Moreover, the expected discrepancies at low frequencies due to non-linearities can be observed in Fig.~\ref{fig:8}, and they are more predominant for the $500~\mathrm{mV}~\mathrm{rms}$ amplitude.

\section{\label{sec5:level1}Conclusions}
A robust experimental setup for performing combined and simultaneous electrochemical and color impedance measurements (EIS and CIS) has been developed. We present results of detailed tests of the technique concerning the signal-to-noise ratio and non-linear effects. To our knowledge, this is the first time that an estimation of the frequency-dependent uncertainties of the CIS measurements has been reported, which is essential for judging the validity of the experimental data in different frequency regions and for comparing the experimental results to theoretical models. Test measurements on an electrochromic $\mathrm{WO}_3$ film demonstrate the versatility of our measurement setup. Further studies using combined EIS and CIS techniques can give relevant insights for developing and testing models for the coloration mechanism of EC systems. Moreover, the technique and theoretical framework presented here can be extended for studying other materials with voltage-modulated optical properties{\textemdash}for example, conducting polymers or other systems for which the coloration mechanism is attributed to interfacial effects.

\begin{acknowledgments}
E. A. Rojas-Gonz\'alez is grateful for the support from the University of Costa Rica.
\end{acknowledgments}

\end{document}